\documentclass[preprint,superscriptaddress,amsmath,amssymb,aps]{revtex4-1}

\usepackage{graphicx}
\usepackage{dcolumn}
\usepackage{bm}

\begin{document}


\title{Out-of-Plane Magnetic Anisotropy in Ordered Ensembles of Fe$_y$N Nanocrystals Embedded in GaN}


\author{A. Navarro-Quezada}
\email{andrea.navarro-quezada@jku.at}
\affiliation{Institute of Semiconductor and Solid-State Physics, Johannes Kepler University Linz, Altenberger Str. 69, 4040 Linz, Austria}

\author{K. Gas}
\affiliation{Institute of Physics, Polish Academy of Sciences, Aleja Lotnikow 32/46, PL-02668 Warsaw, Poland}

\author{T. Truglas}
\affiliation{Christian Doppler Laboratory for Nanoscale Phase Transformations, Center for Surface and Nanoanalytics, Johannes Kepler University Linz, Altenberger Str. 69, 4040 Linz, Austria}

\author{V. Bauernfeind}
\affiliation{Christian Doppler Laboratory for Nanoscale Phase Transformations, Center for Surface and Nanoanalytics, Johannes Kepler University Linz, Altenberger Str. 69, 4040 Linz, Austria}

\author{M. Matzer}
\affiliation{Institute of Semiconductor and Solid-State Physics, Johannes Kepler University Linz, Altenberger Str. 69, 4040 Linz, Austria}

\author{D. Kreil}
\affiliation{Institute of Theoretical Physics, Johannes Kepler University Linz, Altenberger Str. 69, 4040 Linz, Austria}

\author{A. Ney}
\affiliation{Institute of Semiconductor and Solid-State Physics, Johannes Kepler University Linz, Altenberger Str. 69, 4040 Linz, Austria}

\author{H. Groiss}
\affiliation{Christian Doppler Laboratory for Nanoscale Phase Transformations, Center for Surface and Nanoanalytics, Johannes Kepler University Linz, Altenberger Str. 69, 4040 Linz, Austria}

\author{M. Sawicki}
\affiliation{Institute of Physics, Polish Academy of Sciences, Aleja Lotnikow 32/46, PL-02668 Warsaw, Poland}

\author{A. Bonanni}
\affiliation{Institute of Semiconductor and Solid-State Physics, Johannes Kepler University Linz, Altenberger Str. 69, 4040 Linz, Austria}


\date{\today}

\begin{abstract}
Phase-separated semiconductors containing magnetic nanostructures are relevant systems for the realization of high-density recording media. Here, the controlled strain engineering of Ga$\delta$FeN layers with Fe$_y$N embedded nanocrystals (NCs) \textit{via} Al$_x$Ga$_{1-x}$N buffers with different Al concentration $0<x_\mathrm{Al}<41$\% is presented. Through the addition of Al to the buffer, the formation of predominantly  prolate-shaped $\varepsilon$-Fe$_3$N NCs takes place. Already at an Al concentration $x_\mathrm{Al}$\,$\approx$\,5\% the structural properties---phase, shape, orientation---as well as the spatial distribution of the embedded NCs are modified in comparison to those grown on a GaN buffer. Although the magnetic easy axis of the cubic $\gamma$'-Ga$_y$Fe$_{4-y}$N nanocrystals in the layer on the $x_\mathrm{Al} = 0\%$ buffer lies in-plane, the easy axis of the $\varepsilon$-Fe$_3$N NCs in all samples with Al$_x$Ga$_{1-x}$N buffers coincides with the $[0001]$ growth direction, leading to a sizeable out-of-plane magnetic anisotropy and opening wide perspectives for perpendicular recording based on nitride-based magnetic nanocrystals.
\end{abstract}


\maketitle

\section{Introduction}

Iron nitrides (Fe$_y$N) have been widely studied for half a century due to their outstanding physical properties\,\cite{Jack:1952_AC, Eck:1999_JMC, Goelden:2017_JMMM, Coey:1999_JMMM, Leineweber:1999_JAlloyComp, Shirane:1962_PR, Kokado:2006_PRB} and their application in magnetic recording media\,\cite{Coey:1999_JMMM}. Particularly relevant are the high spin polarization and high Curie temperature ($T_\mathrm{C}$) ferromagnetic compounds $\varepsilon$-Fe$_3$N with reported $T_\mathrm{C}\,=\,575$\,K\,\cite{Leineweber:1999_JAlloyComp}, and $\gamma$'-Fe$_4$N with $T_\mathrm{C}\,=\,767$\,K\,\cite{Shirane:1962_PR, Narahara:2009_APL,Kokado:2006_PRB}. Their implementation in combination with GaN into heterostructures is expected to serve for spin injection devices\,\cite{Tao:2010_JCG,Fang:2012_JPD,Kimura:2016_JJAP}. 

In this respect, the controlled fabrication of planar arrays of ferromagnetic $\gamma$'-Ga$_y$Fe$_{4-y}$N nanocrystals (NCs) embedded in a GaN matrix resulting from the epitaxy of Ga$\delta$FeN layers, and whose size, shape and density can be adjusted through the fabrication conditions\,\cite{Navarro:2019_crystals, Navarro:2012_APL}, becomes appealing for the realization of spin injection. The incorporation of Ga ions into the $\gamma$'-Ga$_y$Fe$_{4-y}$N NCs is expected to allow tuning the magnetic properties of the embedded NCs from ferromagnetic to ferrimagnetic\,\cite{Rebaza:2011_JPCC} and weakly antiferromagnetic\,\cite{Houben:2009_ChemMat}, opening wide perspectives for the implementation of these material systems into the field of antiferromagnetic spintronics\,\cite{Jungwirth:2016_NatNano}. The structural, magnetic and transport properties of thin Ga$\delta$FeN layers deposited onto GaN buffers grown on $c$-sapphire (Al$_2$O$_3$) have been already studied in detail\,\cite{Navarro:2012_APL, Grois:2014_Nanotech, Navarro:2019_crystals, Bianco:2018_PCCP, Navarro:2019_PRB}. It was demonstrated that in Ga$\delta$FeN layers, the face-centered cubic $\gamma$'-Ga$_y$Fe$_{4-y}$N nanocrystals have a preferential epitaxial relation $[001]_{\mathrm{NC}}$$\parallel$$[0001]_{\mathrm{GaN}}$ and $\langle\,110\,\rangle_{\mathrm{NC}}\parallel\langle11\bar{1}0\rangle_{\mathrm{GaN}}$, with a minimal fraction of NCs aligned according to $\langle 111 \rangle_{\mathrm{NC}}$$\parallel$$\langle 0001 \rangle_{\mathrm{GaN}}$ and adjusting to the hexagonal symmetry of the matrix. Co-doping with Mn leads to the reduction of the NCs size and to a quenching of the overall superparamagnetic character of the layers\,\cite{Bianco:2018_PCCP}. Recently, in ordered $\gamma$'-Ga$_y$Fe$_{4-y}$N nanocrystal arrays embedded in GaN, the transport of a spin-polarized current at temperatures below 10\,K and an anisotropic magnetoresistance at room-temperature \,\cite{Navarro:2019_PRB} larger than that previously observed for $\gamma$'-Fe$_4$N thin layers\,\cite{Nikolaev:2003_APL}, were observed. 

Further control over these embedded magnetic NCs can be achieved with the modification of their magnetic anisotropy through stress, by incorporating Al into the GaN buffer. The strain energies and piezoelectric effects at the GaN/Al$_x$Ga$_{1-x}$N interface are expected to alter the formation energies and thermodynamic equilibrium conditions of the nanocrystals. In this way, size and shape engineering and the modification of the magnetic anisotropy energy are expected to generate a switchable out-of-plane magnetic anisotropy in the nanocrystals. 

In this work, the effect of strain, induced by adding Al to the GaN buffer---{i.e.,} in Ga$\delta$FeN/Al$_x$Ga$_{1-x}$N ($0 < x_\mathrm{Al} < 41$\%) heterostructures---on the structural and magnetic properties of the Fe-rich nanocrystals embedded in Ga$\delta$FeN thin layers is investigated. It is observed that already 5\% of Al added to the GaN buffer layer modifies not only the structural properties---phase, shape, size and orientation---of the NCs in comparison to those grown on a pure GaN buffer, but it also leads to a sizeable out-of-plane magnetic anisotropy. Through the addition of Al into the buffer layer, additionally to the $\gamma$'-Ga$_y$Fe$_{4-y}$N NCs, the formation of $\varepsilon$-Fe$_3$N NCs is promoted. The crystallographic orientation and the distribution of the two phases in the GaN matrix point at the formation of ordered hexagonal $\varepsilon$-Fe$_3$N NCs elongated along the growth direction as the origin of the observed magnetic anisotropy. 

\section{Experimental details}
The layers considered in this work are grown in a metalorganic vapor phase epitaxy (MOVPE) Aixtron 200X horizontal reactor system (Aixtron, Achen, Germany) on $c$-plane $[0001]$ Al$_2$O$_3$ substrates using trimethylgallium (TMGa), trimethylaluminium (TMAl), ammonia (NH$_3$) and ferrocene (Cp$_2$Fe) as precursors. The 1\,$\mu$m Al$_x$Ga$_{1-x}$N buffers are deposited at 1000$^\circ$C on a 50\,nm low-temperature (540$^\circ$C) Al$_x$Ga$_{1-x}$N nucleation layer annealed at 1000$^\circ$C. The Al concentration $x_\mathrm{Al}$ is varied between 0\% and 41\% over the sample series by adjusting the Ga/Al ratio for the growth of the buffer layer.  

\begin{figure}
	\includegraphics[width=\columnwidth]{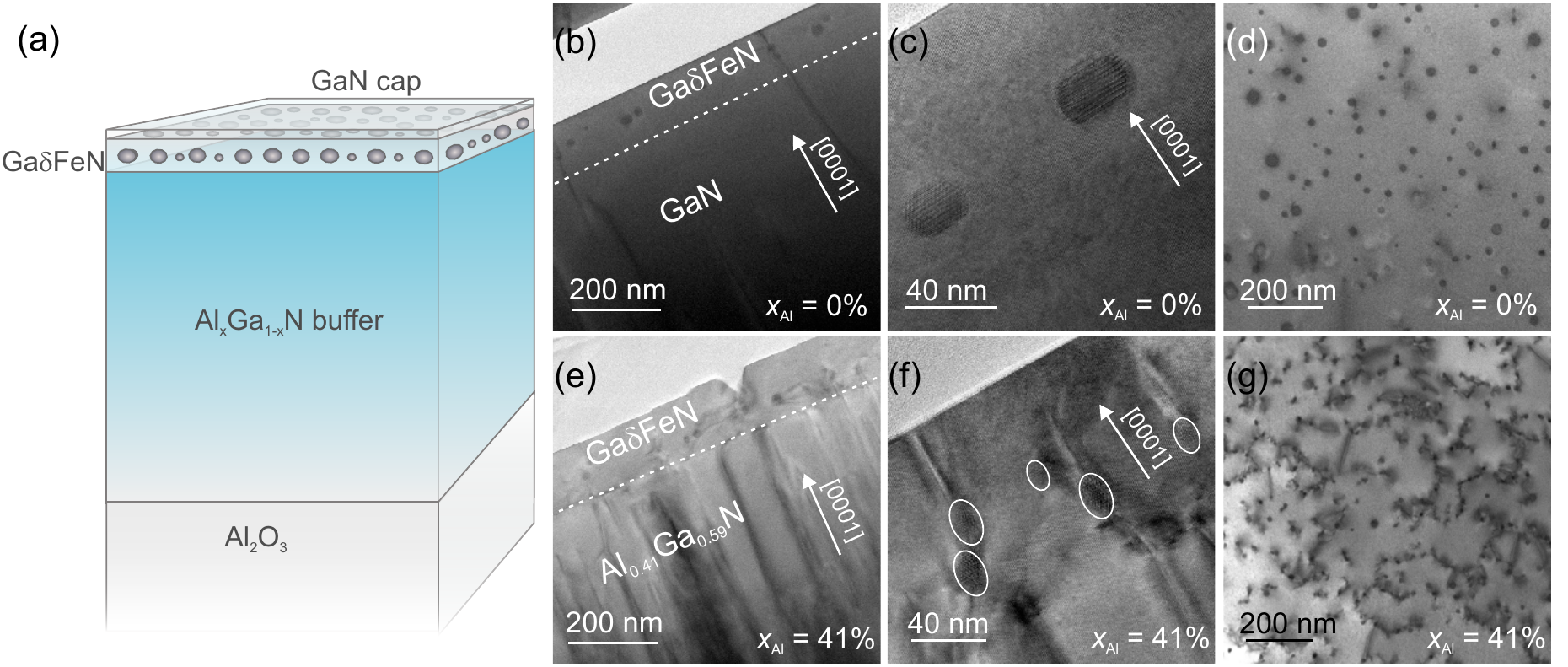}
	\caption{(\textbf{a}): Architecture of the investigated samples. Cross-section TEM micrographs of the samples grown (\textbf{b},\textbf{c}): on GaN, and (\textbf{e},\textbf{f}): on Al$_{0.41}$Ga$_{0.59}$N buffers, showing the embedded nanocrystals distributed in the Ga$\delta$FeN layer. (\textbf{d},\textbf{g}): Plan-view TEM images of the two samples, revealing an increased dislocation network for the layer grown on the Al$_{0.41}$Ga$_{0.59}$N buffer with respect to the layer grown on GaN.}
	\label{fig:struct}
\end{figure}

After deposition of the Al$_x$Ga$_{1-x}$N buffers, a 60\,nm thick Ga$\delta$FeN layer is grown at 810$^\circ$C following the $\delta$-like procedure described in detail in Ref.\,\cite{Navarro:2019_crystals} for Ga$\delta$FeN fabricated onto GaN. The Ga$\delta$FeN layers are covered by a nominally 20\,nm thin GaN capping layer to avoid the segregation to the sample surface of $\alpha$-Fe upon cooling\,\cite{Li:2008_JCG, Navarro:2019_PRB}. A schematic representation of the samples is reproduced in Figure\,\ref{fig:struct}a.

Information on the layers' structure, on $x_\mathrm{Al}$ and on the nanocrystals' phases is obtained by high-resolution X-ray diffraction (HRXRD) carried out in a PANalytical X’Pert Pro Material Research Diffractometer (Malvern Panalytical, Nürnberg, Germany). The measurements have been performed in a configuration that includes a hybrid monochromator equipped with a 0.25$^\circ$ divergence slit, a PixCel detector using 19 channels for detection and a 11.2 mm anti-scatter slit. Rocking-curves acquired along the $[0001]$ growth direction are employed to analyze the overall layer structure and the nanocrystals’ crystallographic phase. From the integral breadth $\beta$ of the (000$l$) symmetric and of the (20$\bar{2}$4) asymmetric diffraction planes, an estimation of the dislocation density in the Al$_x$Ga$_{1-x}$N buffer layers is obtained according to the procedure described by Moram {et al.}\,\cite{Moram:2009_RPP}. Reciprocal space maps (RSM) of the asymmetric (10$\bar{1}$5) diffraction plane allow obtaining directly the in-plane $a$ and out-of-plane $c$ lattice parameters of the Al$_x$Ga$_{1-x}$N buffer and of the Ga$\delta$FeN layers, as well as information on the strain state of the Ga$\delta$FeN layers. The $x_\mathrm{Al}$ is then calculated from the lattice parameters by applying the Vegard’s law\,\cite{Vegard:1921_ZPhysik}.

The structural characterization has been completed by transmission electron microscopy (TEM) imaging using a JEOL JEM-2200FS TEM microscope (Jeol, Tokyo, Japan) operated at 200\,kV in high-resolution imaging (HRTEM) mode. The TEM specimens are prepared in cross-section and plan-view by a conventional procedure including mechanical polishing followed by Ar$^+$ milling. The prepared samples are plasma cleaned before being inserted into the TEM. The elemental analysis is performed \textit{via} energy dispersive X-ray spectroscopy (EDX) of the specimens while measuring the samples in scanning TEM mode (STEM).

The magnetic properties are investigated in a Quantum Design superconducting quantum interference device (SQUID) MPMS-XL magnetometer (Quantum Design, Darmstadt, Germany) equipped with a low field option at magnetic fields $H$ up to 70\,kOe in the temperature range between 2\,K and 400\,K.
The samples are measured in perpendicular and in-plane orientation. The dominant diamagnetic response of the sapphire substrate is compensated by employing a recently developed method for the \textsl{in situ} compensation of the substrate signals in integral magnetometers\,\cite{Gas:2019_MST}.  For the magnetothermal properties, measurements are performed at weak static magnetic fields following the typically employed sequence of measurements: zero-field-cooled (ZFC), field-cooled (FC), and at remanence (TRM). Both ZFC and FC measurements are carried out at $H = 100$\,Oe.
Moreover, since the experimental magnetic signals are in the order of 10$^{-5}$\,emu, all magnetic measurements are carried out by strictly observing an experimental protocol for minute signals\,\cite{Sawicki:2011_SST} elaborated to eliminate artifacts and to overcome limitations associated with integral SQUID magnetometry\,\cite{Pereira:2017_JPDAP}. 

\section{Results and Discussion}
\subsection{Structural properties}{\label{3a}}
The main structural differences between the Ga$\delta$FeN layers grown on GaN and those deposited on the Al$_x$Ga$_{1-x}$N buffers are summarized in Figure\,\ref{fig:struct}, where the overall sample structure, including TEM cross-section and plan-view images for the reference sample ($x_\mathrm{Al}$\,=\,0\%) and for the sample with the highest Al concentration $x_\mathrm{Al}$\,=\,41\% are reported. A comparison between the overview cross-section images presented in Figure \,\ref{fig:struct}b,e reveals a dislocation density in the Al$_{0.41}$Ga$_{0.59}$N buffer larger than the one in GaN, affecting the nanocrystal distribution in the Ga$\delta$FeN overlayer. As a consequence, the NCs are not all localized in one plane like those embedded in the layer grown on GaN, as demonstrated in the TEM micrographs reproduced in Figure\,\ref{fig:struct}c,f. It is further observed that the majority of the NCs in the Ga$\delta$FeN/Al$_{0.41}$Ga$_{0.59}$N sample form at the end of dislocations propagating from the buffer, in contrast to the NCs in the layer grown on GaN, which are embedded in the Ga$\delta$FeN matrix volume. This is visualized in the plan-view images presented in Figure\,\ref{fig:struct}d,g, where NCs with a round-shaped contour, distributed homogeneously in the plane with an average distance of (20--100)\,nm between nanocrystals, are observed. The NCs density increases from (5.0 $\pm$ 0.2)$\times10^{9}$\,NCs/cm$^2$ for the reference sample to (5.0 $\pm$ 0.3)$\times10^{10}$\,NCs/cm$^2$ for the sample grown on the Al$_{0.41}$Ga$_{0.59}$N buffer. Besides an increased NC density, there is a complex dislocation network connecting the NCs observed for the Ga$\delta$FeN layer grown on the Al$_{0.41}$Ga$_{0.59}$N buffer.

\begin{figure}
	\centering
	\includegraphics[width=12 cm]{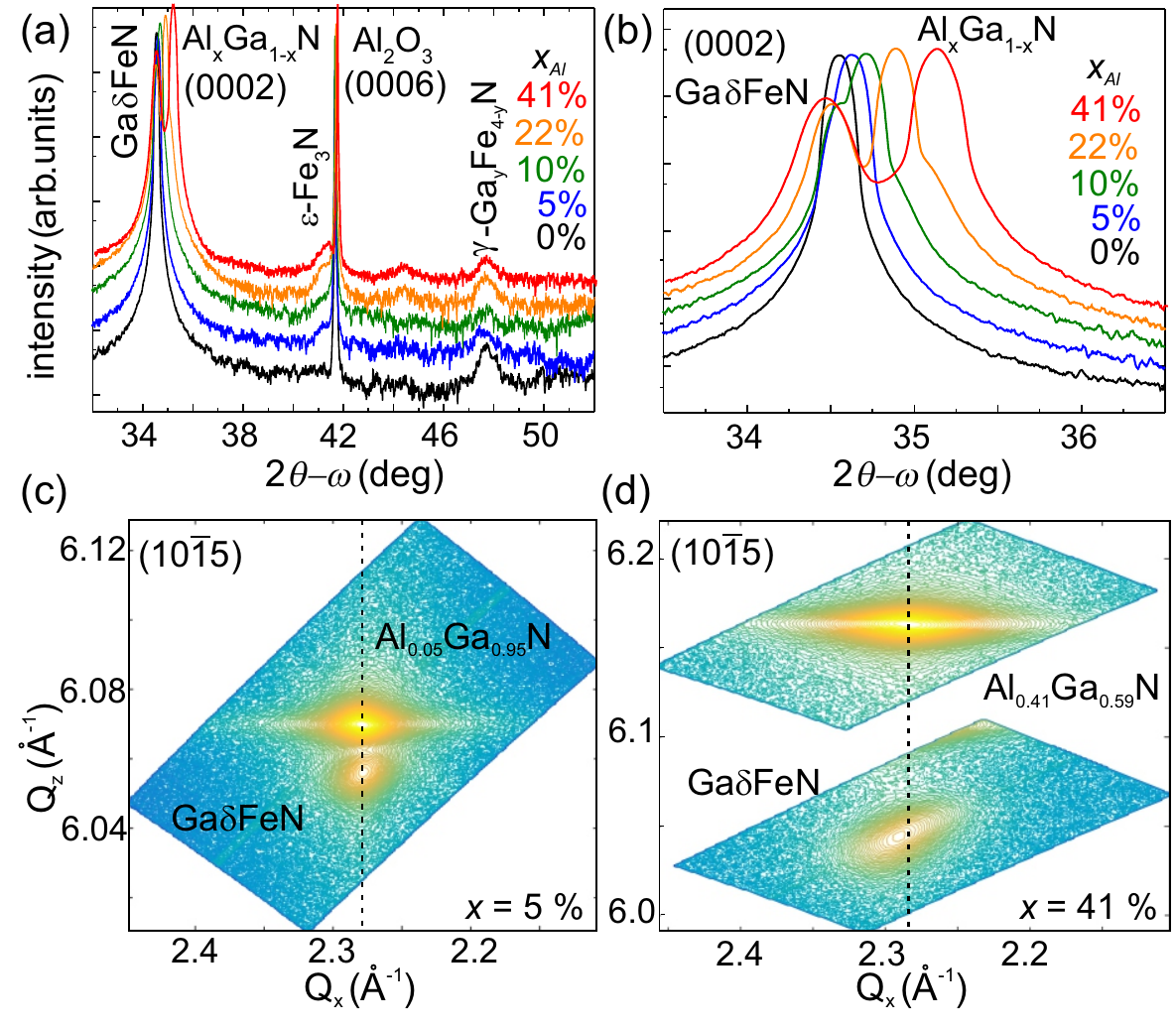}
	\caption{(\textbf{a}) Radial 2$\theta$-$\omega$ scans collected along the $[0001]$ growth direction with the diffraction peaks identified for the Al$_x$Ga$_{1-x}$N buffer, the Ga$\delta$FeN layers and the embedded nanocrystal phases\,\cite{Navarro:2020_crystals}. (\textbf{b})~Close-up of the (0002) diffraction peaks of the Al$_x$Ga$_{1-x}$N buffer and of the Ga$\delta$FeN layers. (\textbf{c},\textbf{d})~Reciprocal space maps of the $(10\bar{1}5)$ diffraction plane for the samples containing 5\% and 41\% Al in the buffer, respectively.}
	\label{fig:xrd}
\end{figure}

The nanocrystal phases are established from the HRXRD 2$\theta$-$\omega$ scans collected along the $[0001]$ growth direction and reported in Figure\,\ref{fig:xrd}a for all samples. Besides the diffraction peaks from the Ga$\delta$FeN layer, from the Al$_x$Ga$_{1-x}$N ($0 < x_\mathrm{Al} < 41$\%) buffer and from the Al$_2$O$_3$ substrate, two additional diffraction peaks located around (41.28$\pm$0.07)$^\circ$ and (47.72$\pm$0.07)$^\circ$ are observed for all samples with Al in the buffer. The first diffraction peak is attributed to the (0002) plane of the hexagonal $\varepsilon$-Fe$_3$N phase, while the second one origins from the (200) plane of the fcc $\gamma$'-Ga$_y$Fe$_{4-y}$N phase. The calculated lattice parameters for the two Fe$_y$N phases are (0.437$\pm$0.002)\,nm and (0.381$\pm$0.002)\,nm, respectively. These values lie in the range of the reported literature values for both phases: the hexagonal $\varepsilon$-Fe$_3$N with $a$\,=\,0.469\,nm and $c$\,=\,0.437\,nm\,\cite{Liapina:2004_AM}, and the fcc $\gamma$'-Ga$_y$Fe$_{4-y}$N with $a$\,=\,0.379\,nm\,\cite{Houben:2009_ChemMat}. For the reference sample, only the $\gamma$'-Ga$_y$Fe$_{4-y}$N phase is observed.

A close-up of the region around the (0002) diffraction peak of the Ga$\delta$FeN overlayer and of the Al$_x$Ga$_{1-x}$N buffer is presented in Figure\,\ref{fig:xrd}b, showing the shift of the buffer peak to higher diffraction angles with increasing Al concentration, pointing at a reduction in the $c$-lattice parameter. The position of the diffraction peak related to the Ga$\delta$FeN thin layer remains unchanged for the buffers with $x_\mathrm{Al}\leq$10\% and shifts to lower angles for increasing Al concentrations, {i.e.,} larger $c$-lattice parameter. This suggests that the Ga$\delta$FeN layer is compressively strained on the Al$_x$Ga$_{1-x}$N buffers.
\begin{table}
	\centering
	\caption{List of investigated samples and their relevant parameters: Al concentration $x_\mathrm{Al}$ in the buffer; R$_{\%}$ degree of relaxation; out-of-plane $\epsilon_{zz}^{\mathrm{GaFeN}}$ and in-plane $\epsilon_{xx}^{\mathrm{GaFeN}}$ strain and $\sigma_{xx}^{\mathrm{GaFeN}}$ stress in the Ga$\delta$FeN thin layer.  The Fe$_y$N nanocrystal phases identified by HRXRD and HRTEM are also listed.}
	\label{tab:tab1}
	\begin{tabular}{c c c c c c }
		\hline \hline
		\textbf{\boldmath{$x_\mathrm{Al}$}} & \textbf{\boldmath{R$_\%$}} & \textbf{\boldmath{$\epsilon_{xx}^{\mathrm{GaFeN}}$}} &\textbf{\boldmath{ $\epsilon_{zz}^{\mathrm{GaFeN}}$}} &  \textbf{\boldmath{$\sigma_{xx}^{\mathrm{GaFeN}}$}} 	& \textbf{\boldmath{Fe$_y$N NCs Phases}} \\
		\hline
		\textbf{(\%)}   & \textbf{ (\%) }         & \textbf{(\%)}  	& \textbf{(\%)}  &  \textbf{(GPa)} 	& 	 \\
		0 & 0 & $-$0.012 & 0.063 & $-$0.564 & $\gamma$'-Ga$_y$Fe$_{4-y}$N \\
		5 & 0 & $-$0.012 & 0.063 & $-$0.564 & $\varepsilon$-Fe$_3$N/$\gamma$'-Ga$_y$Fe$_{4-y}$N \\
		10 & 13 & $-$0.016 & 0.081 & $-$0.706 & $\varepsilon$-Fe$_3$N/$\gamma$'-Ga$_y$Fe$_{4-y}$N \\
		22 & 67 & $-$0.018 & 0.093 & $-$0.847 & $\varepsilon$-Fe$_3$N/$\gamma$'-Ga$_y$Fe$_{4-y}$N \\
		41 & 85 & $-$0.012 & 0.063 & $-$0.564 & $\varepsilon$-Fe$_3$N/$\gamma$'-Ga$_y$Fe$_{4-y}$N \\
		\hline \hline
	\end{tabular}
	
\end{table}

\begin{figure}
	\centering
	\includegraphics[width=8.5 cm]{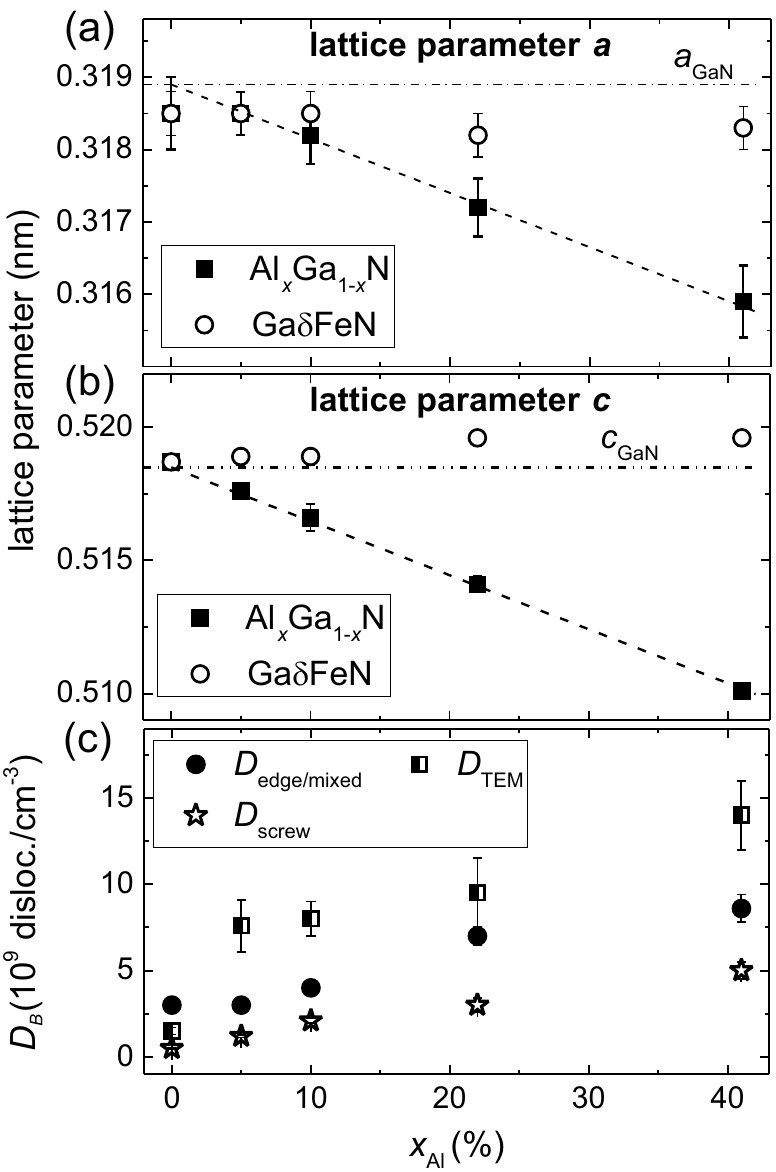}
	\caption{(\textbf{a},\textbf{b}): Lattice parameters $a$ and $c$ of the Al$_x$Ga$_{1-x}$N buffer (full squares) and the Ga$\delta$FeN layers (empty circles) \textit{vs.} $x_\mathrm{Al}$. The dashed line corresponds to the Vegard’s law and the dashed-dotted line indicates the literature values of the lattice parameters $a$ and $c$ for GaN\,\cite{Morkoc:2008}. (\textbf{c}) Dislocation densities---edge-mixed (full circles) and screw (empty stars) – in the Al$_x$Ga$_{1-x}$N buffer layers estimated from XRD and TEM (half-filled squares) as a function of $x_\mathrm{Al}$.}
	\label{fig:xrdparam}
\end{figure}

To analyze the strain state and to obtain the in-plane $a$-lattice parameter, reciprocal space maps at the (10$\bar{1}$5) diffraction plane are acquired. The RSM for the samples with buffers containing 5\% and 41\% of Al are shown in Figure\,\ref{fig:xrd}c and (d), demonstrating that while the Ga$\delta$FeN layer grows fully strained on the Al$_{0.05}$Ga$_{0.95}$N buffer, it is partially relaxed on the Al$_{0.41}$Ga$_{0.59}$N one. The in-plane percentage of relaxation $R_\%$ of the Ga$\delta$FeN thin layer with respect to the buffer is obtained directly from the respective in-plane $d$-lattice spacings as\,\cite{Fewster:1998_TSF}:

\begin{equation}
R_{\%} = \frac{d_{\parallel}^{\mathrm{GaFeN(m)}}-d_{\parallel}^{\mathrm{AlGaN(m)}}}{d_{\parallel}^{\mathrm{GaN(0)}}-d_{\parallel}^{\mathrm{AlGaN(0)}}}\times 100\,,
\end{equation}
\label{eq:one}
where $d_{\parallel}$ refers to the in-plane lattice spacings $d$. The values in the numerator are the measured ones and those in the denominator are the values for free-standing GaN and Al$_x$Ga$_{1-x}$N according to the Vegard’s law. The calculated $R_\%$ values for the samples considered here, are reported in Table\,\ref{tab:tab1}, showing that for $x_\mathrm{Al} <$\,10\%, the Ga$\delta$FeN layers grow fully strained on the buffers and the onset of relaxation occurs at $x_\mathrm{Al}\geq$\,10\%. This is also evident from the lattice parameters presented in Figure\,\ref{fig:xrdparam}a,b as a function of $x_\mathrm{Al}$, where the lattice parameter $a$ for the Ga$\delta$FeN layer is found to deviate from the one of the Al$_x$Ga$_{1-x}$N buffer with $x_\mathrm{Al} >$ 10\%. The dashed lines in Figure\,\ref{fig:xrdparam}a,b give the trend of the Vegard’s law and the dashed-dotted lines indicate the literature values for the lattice parameters for GaN\,\cite{Morkoc:2008}. Although the $c$-lattice parameter for the Ga$\delta$FeN layer is not significantly affected by increasing the Al concentration, $a$ matches the one of the buffer until $x_\mathrm{Al}\approx$10\% and then deviates significantly, confirming the relaxation of the Ga$\delta$FeN thin layer. Considering that the Ga$\delta$FeN thin layer has only biaxial in-plane strain, the strain $\epsilon_{xx}^{\mathrm{GaFeN}}$ and stress $\sigma_{xx}^{\mathrm{GaFeN}}$ tensors are calculated employing a linear interpolation between the value of the Young modulus $E$ and the stiffness constants $C_{ij}$ of GaN ($E$~=~450 GPa, 2$C_{13}/C_{33}$ = 0.509) and AlN ($E$ = 470 GPa, 2$C_{13}$/$C_{33}$ = 0.579)\,\cite{Morkoc:2008}. The values reported in Table\,\ref{tab:tab1} show that independent of the Al concentration, the Ga$\delta$FeN layers are all under a comparable compressive strain.

The (0002) diffraction peak of the Al$_x$Ga$_{1-x}$N buffers presented in Figure\,\ref{fig:xrd}b broadens with increasing Al concentration, pointing at an increment of defects and dislocation density in the buffer layers. In $[0001]$-oriented III-nitride films, three main types of threading dislocations are commonly observed: edge-, mixed- and screw-type. The analysis of the integral breadth of the diffraction peaks originating from the (000$l$) planes allows estimating the density of screw dislocations, while the one in the (20$\bar{2}$4) plane provides information on the density of edge and mixed type dislocations\,\cite{Moram:2009_RPP}. According to Dunn and Koch, the density of dislocations $D_\mathrm{B}$ is given by\,\cite{Dunn:1957_ActaMetall}:
\begin{equation}
D_\mathrm{B} = \frac{\beta^2}{4.35b^2}\,,
\end{equation}
where $\beta$ is the integral breadth and $b$ is the Burgers vector. This equation was previously employed to estimate the dislocation density in GaN thin films\,\cite{Metzger:1998_PMA}. The dislocation densities obtained from HRXRD analysis for all buffer layers as a function of $x_\mathrm{Al}$ are reported in Figure\,\ref{fig:xrdparam}c, where a linear increase is observed reaching values up to four times larger than those of the GaN buffer for both edge-mixed and screw dislocations in the buffer with the highest Al concentration. These results are consistent with the observations from the cross-section and plan-view TEM images shown in Figure\,\ref{fig:struct}. The dislocation density is also estimated from TEM micrographs, yielding larger values for the Al$_x$Ga$_{1-x}$N buffers than those obtained from the XRD analysis, but following the same trend: the greater the concentration of Al in the buffer, the higher the dislocation density.

The increased dislocation density in the Al$_x$Ga$_{1-x}$N buffers with $x_\mathrm{Al}>$\,10\% leads to the relaxation of the Ga$\delta$FeN thin layers. As observed in Figure\,\ref{fig:struct}f, a fraction of the dislocations from the Al$_{0.41}$Ga$_{0.59}$N buffer runs throughout the entire Ga$\delta$FeN layer, promoting the aggregation of Fe along the defects and, therefore, the preferential formation of nanocrystals. Interestingly, the nanocrystals stabilized at the dislocations are predominantly elongated along the $[0001]$ growth direction.

\begin{figure}
	\center
	\includegraphics[width=14 cm]{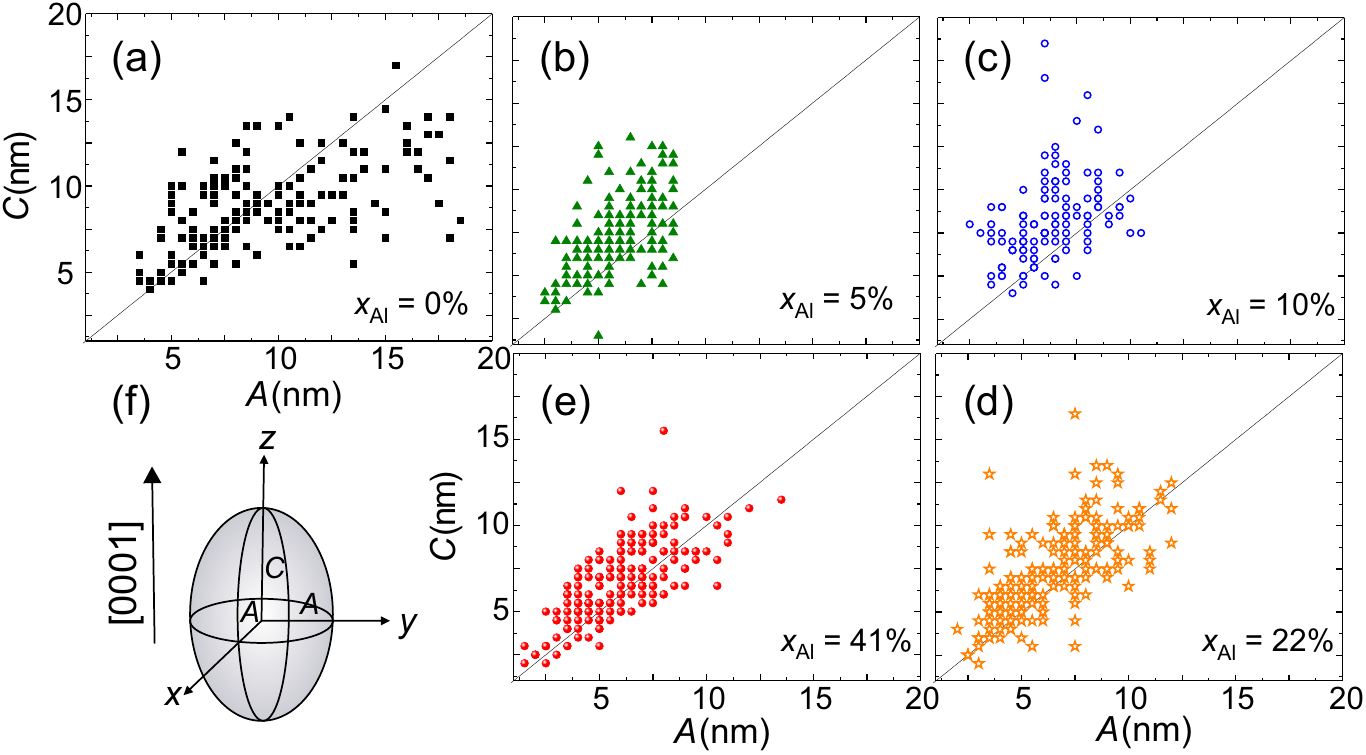}
	\caption{Size distribution of 200 NCs measured in cross-section HRTEM for $x_\mathrm{Al}$ in the buffers equal to: (\textbf{a}) 0\%, (\textbf{b}) 5\%, (\textbf{c}) 10\%, (\textbf{d}) 22\%, and (\textbf{e}) 41\%. The dimensions $A$ and $C$ correspond to the schematic representation depicted in (\textbf{f}) and correspond to half the size perpendicular and parallel to the $[0001]$ growth direction, respectively.}
	\label{fig:temparam}
\end{figure}

A more detailed analysis of the NCs sizes is performed on cross-section and plan-view TEM images. The size of the NCs is determined with an accuracy of $\pm$0.5\,nm by measuring the size of the areas where Moir\'{e} patterns are visible with the Fiji software\,\cite{Schindelin:2012_Natmet}. The results are presented in Figure\,\ref{fig:temparam}a--e, where the size distribution of 200 measured NCs per sample is reported. For this evaluation, the NCs are treated as ellipsoids according to the schematic representation in Figure\,\ref{fig:temparam}f with dimensions perpendicular ($A$) and parallel ($C$) to the $[0001]$ growth direction for the different $x_\mathrm{Al}$ in the buffers. The solid line marks the aspect ratio (AR) equal to 1, {i.e.,} $A=C$. From the size distributions presented in Figure\,\ref{fig:temparam}, it is seen that the size of the NCs in the reference sample has a broader distribution and particularly a larger in-plane $A$ than in the samples grown on the Al$_x$Ga$_{1-x}$N buffers. Although the size of the NCs in the reference sample tends to lie on or below the solid line, indicating an AR$\leq$1 and an oblate shape of the NCs---with their $y$-axis elongated in the plane of the layer---the size of the NCs in the layers grown on the Al$_x$Ga$_{1-x}$N buffers lies above the solid line, {i.e.,} with an AR>1, pointing at prolate NCs elongated along the $[0001]$ growth direction. From the measured dimensions of the NCs, the average sizes parallel and perpendicular to the growth direction $[0001]$ are estimated, confirming the decrease in the size perpendicular to the growth direction for the nanocrystals embedded in the Ga$\delta$FeN layers grown on the Al$_x$Ga$_{1-x}$N buffers.

\begin{figure}
	\centering
	\includegraphics[width=14 cm]{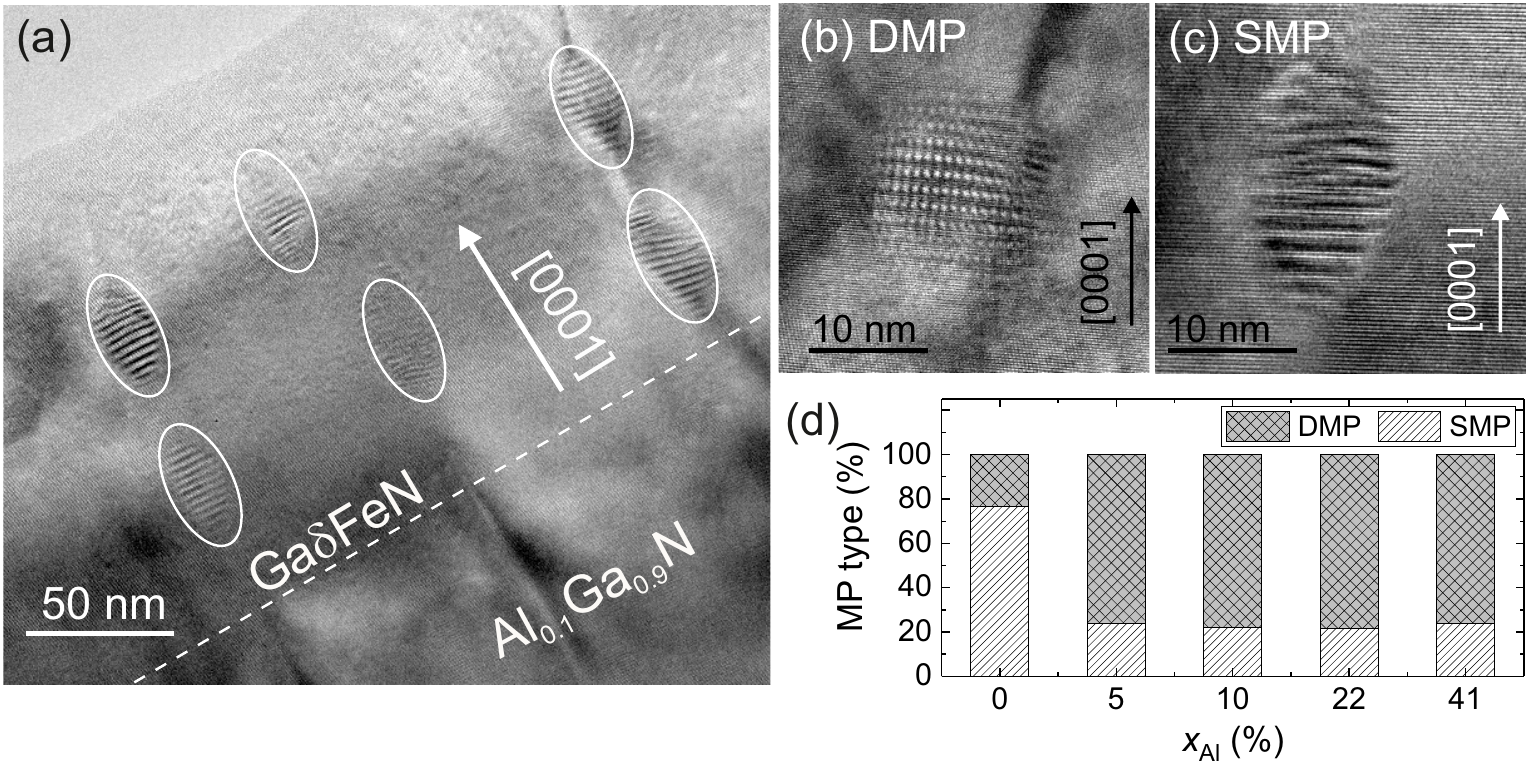}
	\caption{(\textbf{a}) Cross-section HRTEM image showing the distribution in pairs of prolate NCs along dislocations in the Ga$\delta$FeN/Al$_{0.1}$Ga$_{0.9}$N sample. (\textbf{b},\textbf{c})  HRTEM images of nanocrystals with double and single Moir\'{e}-patterns, respectively. (\textbf{d})  Fraction of NCs displaying SMP and DMP as a function of $x_\mathrm{Al}$.}
	\label{fig:mp}
\end{figure}

Furthermore, it is found that in all samples the nanocrystals located at dislocation sites are predominantly prolate. This suggests that the increase in dislocation density for the layers grown on the Al$_x$Ga$_{1-x}$N buffers promotes the formation of prolate NCs, which are mostly arranged in pairs aligned along dislocations, as shown in Figure\,\ref{fig:mp}a. In contrast, the oblate NCs are all located at the same depth in the layers.

In addition to providing the size and phase, the characterization of the Moir\'{e} patterns (MPs) observed in the HRTEM micrographs yields further relevant information about the embedded NCs. The origin of MPs in general is the result of the overlap of two lattices with equal spacings that are rotated with respect to each other, or of the superposition of lattices with slightly different spacings. This leads to a pattern with Moir\'{e} fringe spacings with either single periodicity (line pattern) or double periodicity (grid-like pattern). Exemplary NCs showing a double and a single MP are presented in Figure\,\ref{fig:mp}b,c, respectively. The Moir\'{e} fringe spacings depend on the two underlying crystal structures, on their orientation relationship, and on the lattice strain. The fraction of nanocrystals displaying single MP (SMP) and double MPs (DMP) is shown in Figure\,\ref{fig:mp}d. Up to 78\% of the NCs exhibit single MPs and 22\% produce double MPs in the reference Ga$\delta$FeN grown on GaN buffer, while for the films grown on the Al$_x$Ga$_{1-x}$N buffers this tendency is inverted. The double MP pattern is an indication of an in-plane misorientation of the NCs, which is related to the enhanced dislocation density in the underlying buffer layers and to the formation of the NCs along the dislocations, leading to slight distortions and strain within the GaN matrix.

\begin{figure}
	\centering
	\includegraphics[width=13 cm]{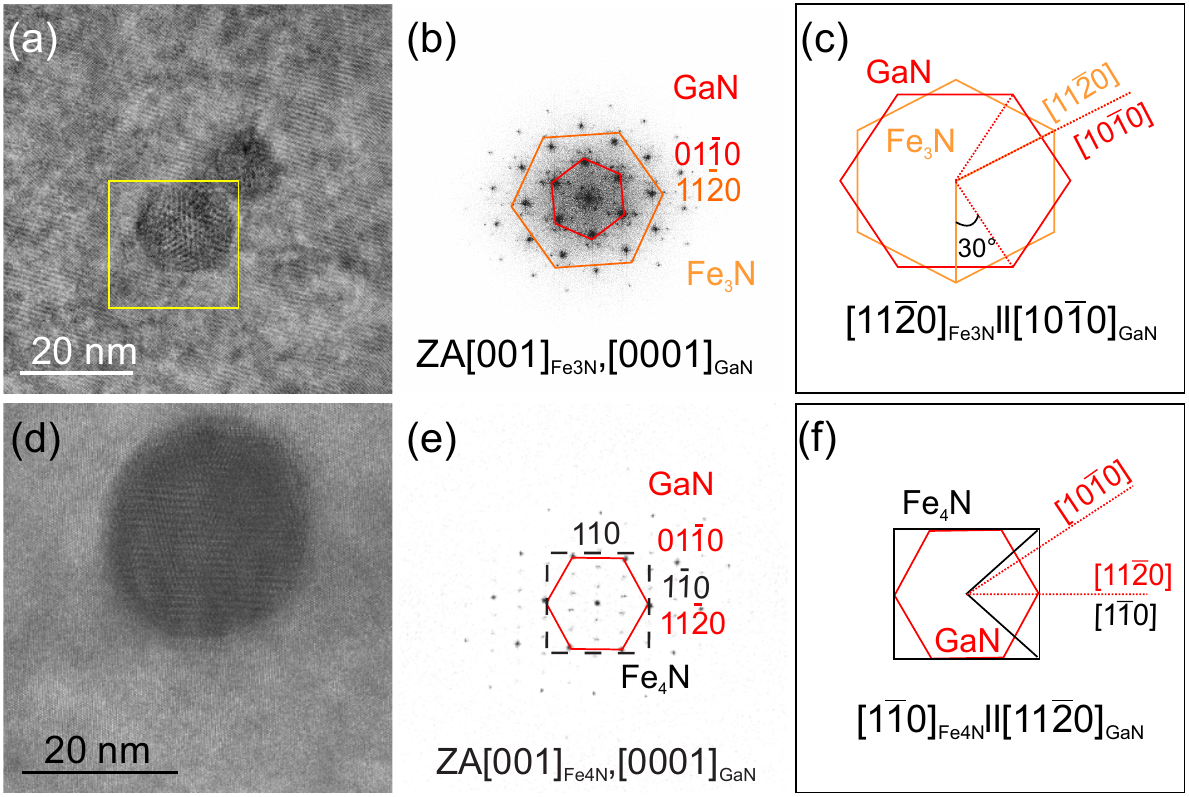}
	\caption{Plan-view HRTEM images of exemplary Fe$_y$N nanocrystals embedded in a Ga$\delta$FeN layer grown on (\textbf{a})  an Al$_{0.1}$Ga$_{0.9}$N buffer, and (\textbf{d})  GaN. (\textbf{b},\textbf{e})  FFT of the images presented in (\textbf{a},\textbf{d}), respectively, showing the epitaxial orientation of the NCs with respect to the GaN matrix. The FFT in (\textbf{c}) corresponds to the NCs marked by the square in (\textbf{a}). (\textbf{c},\textbf{f}): Schematic representation of the epitaxial relation in (\textbf{b},\textbf{e}).}
	\label{fig:tem2}
\end{figure}

The Fe$_y$N phases identified in the HRXRD spectra depicted in Figure\,\ref{fig:xrd}a are confirmed by HRTEM analysis. In HRTEM micrographs showing NCs, the regions of interests are Fourier transformed by Fast Fourier Transformation (FFT) using the Gatan Digital Micrograph (Gatan Inc.) software. Micrographs of two NCs are shown in Figure\,\ref{fig:tem2}a,d along with the corresponding FFTs in Figure\,\ref{fig:tem2}b,e. The FFT images are used to determine the lattice parameters by measuring the spacings in the two directions of the diffraction pattern. To identify the NCs orientation with respect to the GaN matrix, a~comparison with the diffraction patterns simulated by the JEMS software is performed\,\cite{Stadelmann:1987_Umicroscopy}. Employing this procedure, the investigated NC in Figure\,\ref{fig:tem2}a is identified as $\varepsilon$-Fe$_3$N oriented along the zone axis (ZA) $[110]_{\mathrm{NC}}$, which is parallel to the ZA $[210]_{\mathrm{GaN}}$, and therefore corresponds to an epitaxial relation $[11\bar{2}0]_{\mathrm{NC}}$$\parallel$$[10\bar{1}0]_{\mathrm{GaN}}$. A schematic representation of the epitaxial relation is sketched in Figure\,\ref{fig:tem2}c, showing that the NC is 30$^\circ$ rotated with respect to the crystallographic axis of GaN, but parallel to the one of the sapphire substrate, similarly to the fcc NCs studied in Ga$\delta$FeN/GaN layers\,\cite{Navarro:2012_APL}. The above procedure is applied to the NCs found in the reference sample and reproduced in Figure\,\ref{fig:tem2}d, revealing the epitaxial relation $[110]_{\mathrm{NC}}$$\parallel$$[11\bar{2}0]_{\mathrm{GaN}}$ presented in Figure\,\ref{fig:tem2}f  and previously reported for $\gamma$'-Ga$_y$Fe$_{4-y}$N NCs in Ga$\delta$FeN layers grown on GaN\,\cite{Navarro:2012_APL}. The majority of the NCs found in the Ga$\delta$FeN layers grown on the Al$_x$Ga$_{1-x}$N buffers are identified as the hexagonal $\varepsilon$-Fe$_3$N phase, while those in the reference sample are associated with the cubic $\gamma$'-Ga$_y$Fe$_{4-y}$N phase oriented preferentially as $[001]_{\mathrm{NC}}$$\parallel$$[0001]_{\mathrm{GaN}}$, in agreement with the results from the XRD spectra presented in Figure\,\ref{fig:xrd}a.
From elemental composition analysis \textit{via} EDX line-scans, the presence of Al in the Ga$\delta$FeN layers is ruled out as shown in Fig.\,S1 of the Suplemental Material.  

\subsection{Magnetic properties}

In the previous section it has been demonstrated that the basic structural characteristics of the NCs change considerably with the incorporation of Al into the buffer layer. To shed light onto how the magnetic characteristics of the layers are modified by these structural changes, a comparative analysis of the magnetic properties of the reference Ga$\delta$FeN/GaN and the Ga$\delta$FeN/Al$_{0.1}$Ga$_{0.9}$N samples is performed. As indicated in Table\,\ref{tab:tab1} and depicted in Figure\,\ref{fig:xrd}, the former contains mostly $\gamma$'-Ga$_y$Fe$_{4-y}$N NCs, which are characterized by a balanced distribution of prolate and oblate shapes, whereas in the latter, prolate $\varepsilon$-Fe$_3$N NCs prevail over the $\gamma$'-Ga$_y$Fe$_{4-y}$N ones.

The formation of the Fe-rich NCs in GaN is the direct consequence of the solubility limit of Fe in GaN being ($1.8 \times 10^{20}$)\,cm$^{-3}$ or  0.4\% at the growth conditions considered here\,\cite{Przybylinska:2006_MSEB,Bonanni:2007_PRB, Navarro:2010_PRB}.
Therefore, when the doping level exceeds this concentration, the Fe ions are found both in Ga substitutional sites as Fe$^{3+}$ and in the phase-separated NCs. The Fe-rich NCs form disperse ensembles of large ferromagnetic  macrospins with specific size and shape distributions. In the absence of mobile carriers, the randomly distributed Fe$^{3+}$ ions, despite their high spin state ($L=0, S=5/2$), do not interact in the relevant temperature range between 2\,K and 400\,K and exhibit paramagnetic properties. Due to the high diffusivity of transition metal ions in GaN, these paramagnetic ions are found diffusing a few hundreds of nanometres below the Fe-$\delta$-doped layer\,\cite{Jakiela:2019_JAC}. This substantially increases the total amount of the dilute Fe$^{3+}$, making the intensity of the paramagnetic signal at low temperatures comparable to the one of the ferromagnetic NCs. Therefore, a dedicated experimental approach is required to distinguish between the two contributions.

\begin{figure}
	\centering
	\includegraphics[width=15 cm]{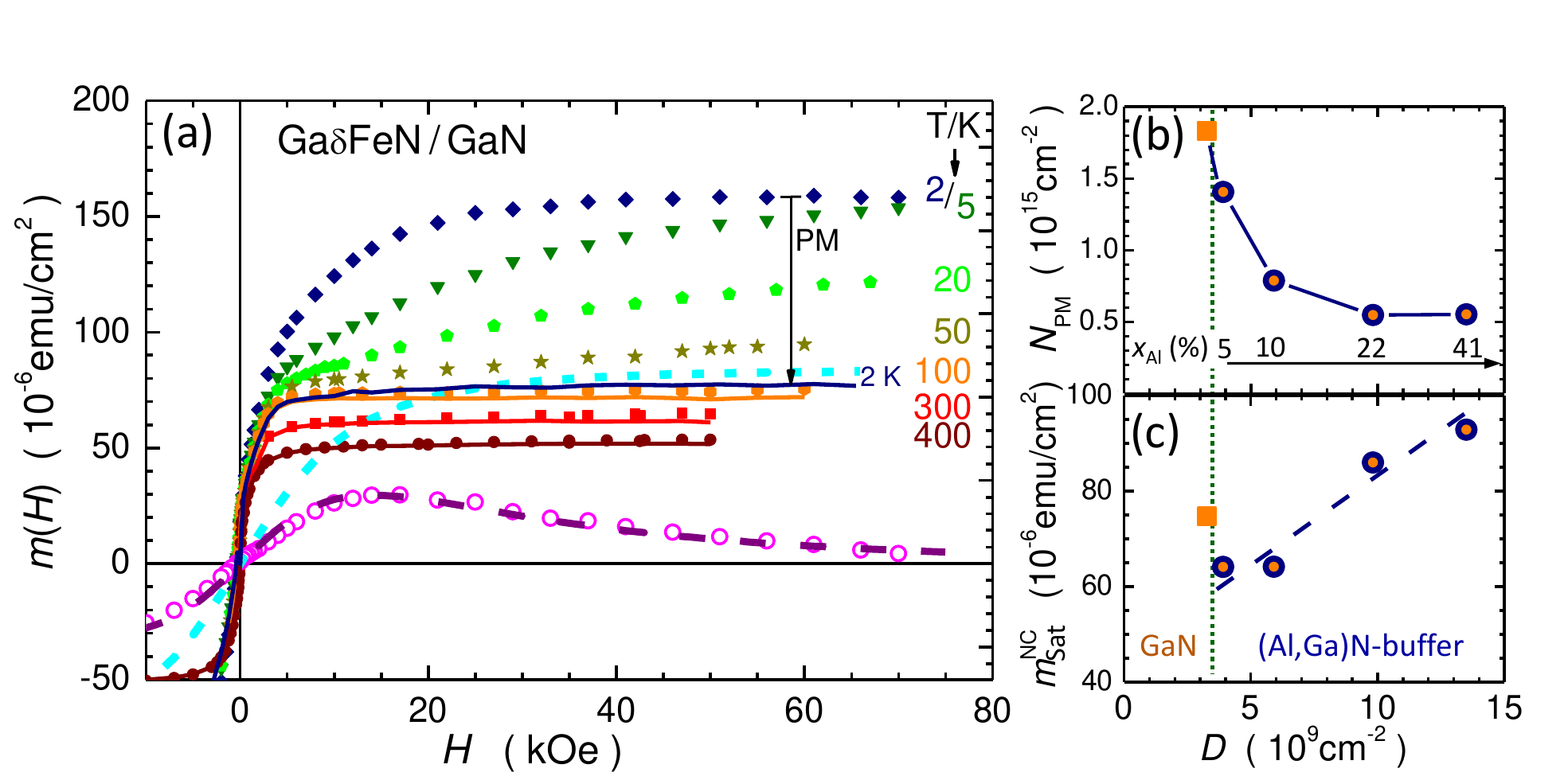}
	\caption{(\textbf{a})  (Solid symbols) Isothermal magnetization curves of the reference Ga$\delta$FeN/GaN structure at selected temperatures. The open circles denote the difference $\Delta m(H)$, whereas the dashed line corresponds to the calculated difference of the respective Brillouin functions calculated for the paramagnetic Fe$^{3+}$ ions with $N_{\mathrm{PM}}\,=\,(1.8 \times 10^{15})$\,cm$^{-2}$. The solid lines mark the resulting magnitudes of $m_{\mathrm{NC}}(H)$ of the NCs, after subtracting the paramagnetic component. The solid down--arrow indicates the degree of the reduction of $m(H)$ due to the subtraction of the paramagnetic contribution. (\textbf{b},\textbf{c})  $N_{\mathrm{PM}}$ and $m_{\mathrm{NC}}^{sat}$ plotted as a function of total dislocation density $D$. The squares represent the reference Ga$\delta$FeN/GaN structure, the circles mark data for the layers grown on the Al$_x$Ga$_{1-x}$N buffers. The corresponding concentration of Al in the Al$_x$Ga$_{1-x}$N buffers is indicated in panel (\textbf{b}). Dashed lines in panels (\textbf{b},\textbf{c}) are guide to the eye.}
	\label{fig:M1}
\end{figure}

The isothermal magnetization curves with the magnetic moment as a function of the applied magnetic field $m(H)$ for the reference sample ($x_\mathrm{Al}\,=\,0$\%) are plotted for selected temperatures (solid symbols) in Figure\,\ref{fig:M1}. 
As mentioned, the bare magnetic signal consists of two distinct contributions. At~temperatures above 50\,K, the fast saturating response resembling a Langevin's $L(H)$ function at weak fields is attributed to the ferromagnetic NCs. However, the lack of a systematic $T$-dependency satisfying the $H/T$ scaling\,\cite{Bean:1956_JAP} and the presence of a weak magnetic hysteresis indicate that the majority of the NCs is not in thermal equilibrium and their magnetic response is affected by the presence of energy barriers and governed by their distribution. At temperatures below 50\,K, the $m(H)$ gains in strength and a slowly saturating contribution originating from the non-interacting Fe$^{3+}$ ions retaining their own magnetic moment dominates\,\cite{Bonanni:2007_PRB,Navarro:2010_PRB,Nielsen:2012_PRB, Sawicki:2013_PRB}.

The paramagnetism of the Fe$^{3+}$ ions is described by the Brillouin function $B_{S}$ for $S = J = 5/2$~\cite{Przybylinska:2006_MSEB,Pacuski:2008_PRL,Malguth:2008_PSSB}, and the experimentally established difference $\Delta m(H)$ between $m(H)$ at,  {e.g.}, 2\,K and 5\,K permits the quantification of the ions' contribution by fitting $\Delta B_{S}(H,\Delta T)=B_{S}(H,2\,\mathrm{K})-B_{S}(H,5\,\mathrm{K})$ to $\Delta m(H)$ with the procedure described in detail in Ref.\,\cite{Navarro:2010_PRB}.
The open circles in Figure\,\ref{fig:M1} represent the experimental difference $\Delta m(H)$ between $m(H)$ at 2\,K and 5\,K, whereas the dotted line follows the magnitude of the expected change $\Delta B_{5/2}(H,\Delta T)$ corresponding to several ions $N_{\mathrm{PM}} = (1.8 \times 10^{15})$ cm$^{-2}$. The dashed line indicates the magnitude of the paramagnetic contribution corresponding to $N_{\mathrm{PM}}$ at 2\,K.

Having established $N_{\mathrm{PM}}$ in each of the investigated structures, the paramagnetic contribution $m_{\mathrm{PM}}(H)=g\mu_{\mathrm{B}}S N_{\mathrm{PM}} B_{5/2}(H,T)$---where $g$ is the g-factor and $\mu_\mathrm{B}$ the Bohr magneton---is calculated and subtracted from the experimental data to obtain the magnitude $m_{\mathrm{NC}}(H,T)$ of the magnetization corresponding to the NCs. The results are indicated by solid lines in Figure\,\ref{fig:M1}. It is worth noting that  $m_{\mathrm{NC}}(H,T)$ saturates at all investigated temperatures for $H\,\geq\,10$\,kOe, confirming the ferromagnetic order within the NCs. The evolution of $N_{\mathrm{PM}}$ and $m_{\mathrm{NC}}$ as a function of the dislocation density is presented in Figure\,\ref{fig:M1}b,c, respectively. The former decreases, whereas the latter increases with the dislocation density, suggesting that the dislocations originating at the sapphire/Al$_x$Ga$_{1-x}$N interface serve as preferential sites for the aggregation of the Fe ions. This is substantiated by the fact that the magnitude of $N_{\mathrm{PM}}$ in the reference structure and related solely to the layer nominally containing Fe, {i.e.,} (60--100)\,nm, corresponds to ($4\,\times 10^{20}$)\,cm$^{-3}$ or $\simeq 1$\% of Fe ions, largely exceeding the Fe solubility limit in GaN. Thus, the Fe$^{3+}$ ions are distributed across the entire depth in the structure of the reference sample, whereas in the layers grown on the Al$_x$Ga$_{1-x}$N buffers a significant fraction of the Fe ions migrates to the dislocations, where they aggregate into the hexagonal $\varepsilon$-Fe$_3$N NCs. Since the dislocation density is found to correlate with the Al content in the buffer, as presented in Figure\,\ref{fig:xrdparam}c, the Al content in the Al$_x$Ga$_{1-x}$N buffer is instrumental to control both the substitutional Fe atoms concentration and the strength of the ferromagnetic signatures related to the NCs.

\begin{figure}
	\centering
	\includegraphics[width=13 cm]{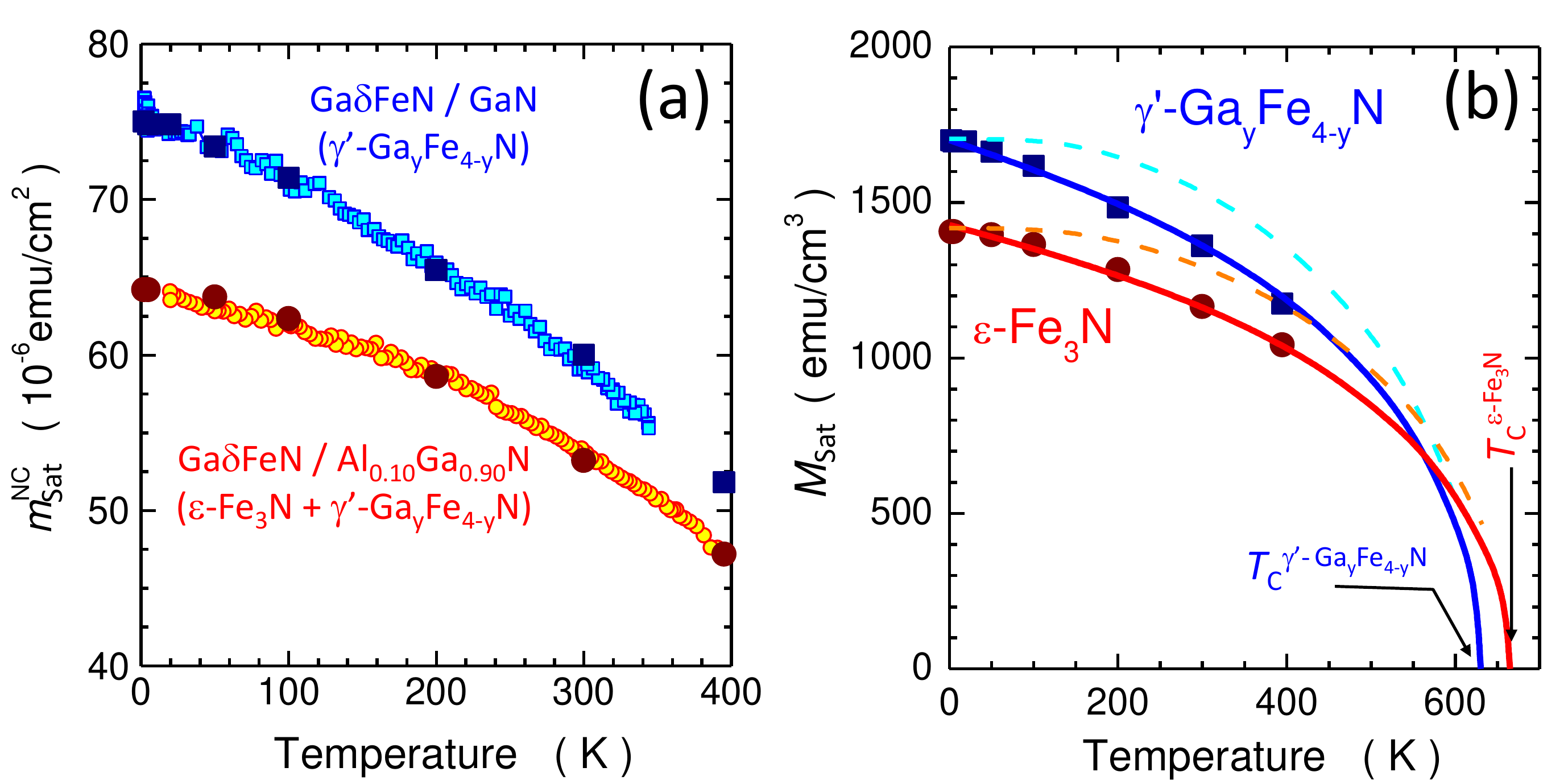}
	\caption{(\textbf{a})  Comparison of the temperature dependence of $m_{\mathrm{NC}}^{\mathrm{sat}}(T)$ in the studied Ga$\delta$FeN layers grown on a GaN buffer (squares) and grown on a Al$_{0.1}$Ga$_{0.9}$N buffer (circles). Solid symbols: $m_{\mathrm{NC}}^{\mathrm{sat}}$ inferred from the $m_{\mathrm{NC}}(H)$ isotherms. Open symbols: direct continuous sweeping of $T$ at $H=20$\,kOe. (\textbf{b}) Temperature dependence of the saturation magnetization $M_{\mathrm{Sat}}$ of the two Fe$_y$N compounds formed due to the epitaxy of the Ga$\delta$FeN layers. The solid lines mark two classical Langevin functions $L(T)$ rescaled to follow the corresponding experimental result for $2\,$K\,$< T < 400$\,K. The dashed lines are Brillouin functions $B_{5/2}(T)$ rescaled to reproduce the corresponding magnitudes of $m_{\mathrm{NC}}^{\mathrm{sat}}(0)$ and $T_{\mathrm{C}}$. }
	\label{fig:MTemp}
\end{figure}

The temperature dependence of the saturation magnetization $m_{\mathrm{NC}}^{\mathrm{sat}}(T)$ of the ferromagnetic signal specific to the NCs for the layer grown on the Al$_{0.1}$Ga$_{0.9}$N buffer (circles) and for the reference one (squares) is reproduced in Figure\,\ref{fig:MTemp}.
These dependencies are established upon performing a $m(H)$ analysis similar to the one exemplified in Figure\,\ref{fig:M1} (solid symbols), as well as from direct continuous sweeping of $T$ at $H=20$\,kOe (open symbols). This allows quantifying the temperature dependence of the saturation magnetization $M_{\mathrm{sat}}$ of the $\gamma$'-Ga$_y$Fe$_{4-y}$N and $\varepsilon$-Fe$_3$N  present in the structures.

To quantify the magnetization of the NCs, their average volume is estimated from the size distribution shown in Figure\,\ref{fig:temparam} and the average densities established from TEM by taking into account that (50-70)\% of the prolate NCs in the Ga$\delta$FeN/Al$_x$Ga$_{1-x}$N structures grow in pairs along the dislocations, as shown in Figure\,\ref{fig:mp}a. The estimated values of the NCs magnetization are $(1700\,\pm\,200)$\,emu/cm$^3$ for the NCs in the reference sample containing $\gamma$'-Ga$_y$Fe$_{4-y}$N NCs, and $(1400\,\pm\,900)$\,emu/cm$^3$  for the NCs present in the Ga$\delta$FeN/Al$_{0.1}$Ga$_{0.9}$N structure, where about 80\% of the NCs are $\varepsilon$-Fe$_3$N and 20\% are $\gamma$'-Ga$_y$Fe$_{4-y}$N. These values are consistent with those estimated from ferromagnetic resonance measurements\,\cite{Grois:2014_Nanotech}, shown in Fig.\,S2 of the Suplemental Material, and in good agreement with the respective ranges of $M_{\mathrm{sat}}$ reported in the literature for these compounds. 
For $\gamma$'-Fe$_4$N, the $M_{\mathrm{sat}}$ ranges between 1500\,emu/cm$^3$ and 2000\,emu/cm$^3$ \cite{Eck:1999_JMC,Xiao:1994_APL,Atiq:2008_APL,Dirba:2015_JMMM}, so that the values obtained for the $\gamma$'-Ga$_y$Fe$_{4-y}$N NCs considered here point at high crystallinity and low dilution by Ga, {i.e.,} $(y \ll 1)$. 
For the layer grown on the Al$_{0.1}$Ga$_{0.9}$N buffer the $M_{\mathrm{sat}}$ established, taking into account a 20\% contribution of $\gamma$'-Ga$_y$Fe$_{4-y}$N NCs, yields a corrected value of $M_{\mathrm{sat}} = (1300 \pm 900)$\,emu/cm$^3$ for the $\varepsilon$-Fe$_3$N NCs, consistent with previous studies\,\cite{Leineweber:1999_JAlloyComp,Eck:1999_JMC,Bhattacharyya:2010_CPL,Robbins:1964_JPCS,Wu:2004_JALCOM,Siberchicot:1993_IJMPB,Yamaguchi:2007_JCG,Zieschang:2017_CM,Mamiya:2002_TMSJ}.

The resulting magnitudes of $M_{\mathrm{Sat}}(T)$ for both compounds are represented as solid symbols in Figure~\ref{fig:MTemp}b. The experimental trends of $M_{\mathrm{Sat}}(T)$ for both Fe$_y$N compounds are compared with the spontaneous magnetization calculated as a function of $T$ based on the molecular field theory in the classical limit and with the Langevin function $L(T)$, {i.e.,} corresponding to a large magnetic moment of the NCs $J=S \rightarrow \infty$ (solid lines). It is observed that the low-$T$ fast drop of $m_{\mathrm{FM}}(T)$ starting at $T\approx50$\,K, is indeed well captured by $L(T)$, and could not be reproduced by a Brillouin function. For comparison, the $B_{5/2}(T)$ functions are added to Figure\,\ref{fig:MTemp}b as dashed lines. The $L(T)$ is then extrapolated to assess the $T_{\mathrm{C}}$ of the NCs in each sample.

In the reference sample containing mostly $\gamma$'-Ga$_y$Fe$_{4-y}$N NCs a $T_{\mathrm{C}}\,=(\,630\,\pm\,30)$\,K is found, {i.e.,} about 100\,K lower than the values reported for Ga-free $\gamma$'-Fe$_{4}$N of $T_{\mathrm{C}}\,=\,716$\,K\,\cite{Dirba:2015_JMMM} and 767\,K  \cite{Shirane:1962_PR}. This is attributed to a partial replacement of the Fe ions by Ga, which leads to a magnetic dilution and randomization of spins breaking down the ferromagnetic order\,\cite{Houben:2009_ChemMat, Burghaus:2011_JSSC}. However, the Ga incorporation is minimal, since the ternary GaFe$_3$N is weakly antiferromagnetic\,\cite{Houben:2009_ChemMat}. The same extrapolation method yields $T_{\mathrm{C}}\,=(670\,\pm\,30)$\,K for the layer grown on the Al$_{0.1}$Ga$_{0.9}$N buffer, which contains predominantly $\varepsilon$-Fe$_3$N NCs and a limited amount of $\gamma$'-Ga$_y$Fe$_{4-y}$N. No quantitative conclusion about the $T_{\mathrm{C}}$ of $\varepsilon$-Fe$_3$N NCs can be made, nevertheless it can be stated that its value is significantly greater than the previously reported 575\,K\,\cite{Leineweber:1999_JAlloyComp} and (500--525)\,K\,\cite{Zieschang:2017_CM,Mukasyan:2019_IC}. This result is relevant, since despite the high potential of $\varepsilon$-Fe$_3$N for spintronics\,\cite{Leineweber:1999_JAlloyComp}, the technological development of this material has been limited by its high chemical reactivity and by challenges in obtaining the required stoichiometry\,\cite{Gajbhiye:2008_MCP}. The magnitude reported here for $\varepsilon$-Fe$_3$N NCs points, on the other hand, to the possibility of stabilizing, in a controlled fashion, relevant Fe$_y$N nanostructures in a GaN matrix. 

\begin{figure}
	\centering
	\includegraphics[width=15 cm]{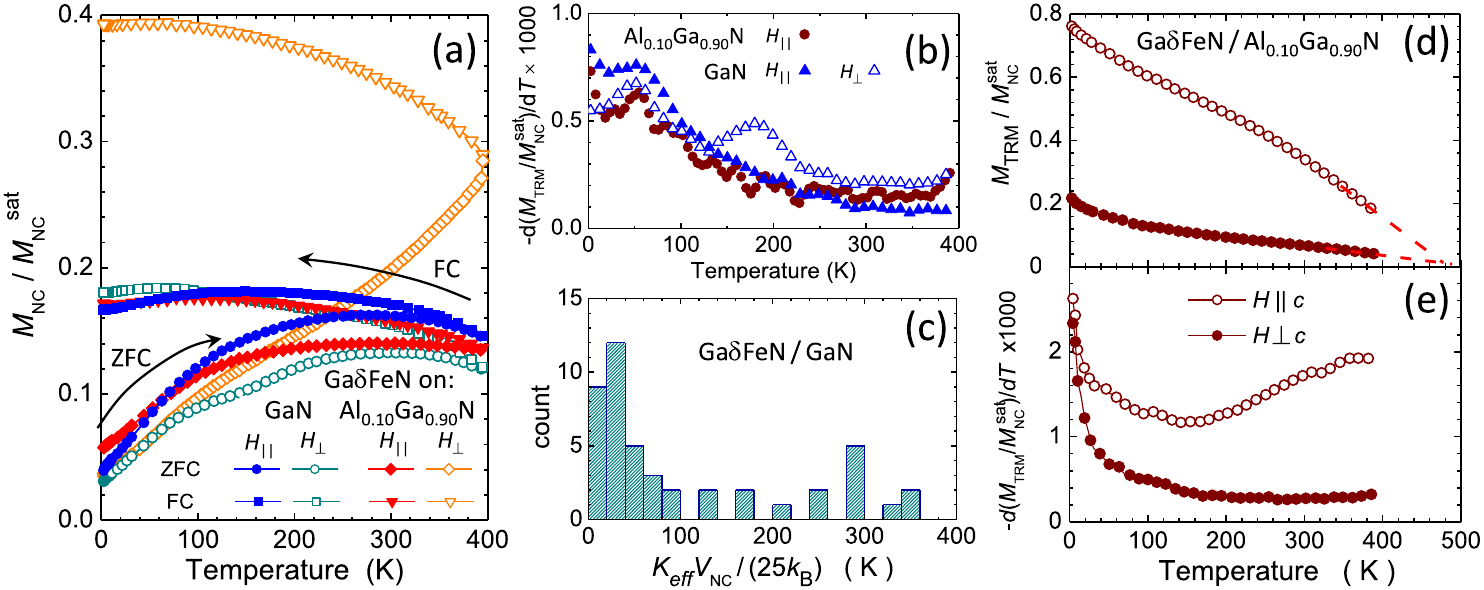}
	\caption{(\textbf{a},\textbf{b})  ZFC, FC and the calculated temperature derivative of the thermoremanence magnetization (TRM): $-d(M_{\mathrm{FC}} - M_{\mathrm{ZFC}})dT$ in the studied Ga$\delta$FeN structures grown either on GaN or on the Al$_{0.1}$Ga$_{0.9}$N buffer. (\textbf{c})  Superparamagnetic limit  distribution in the Ga$\delta$FeN/GaN  structure calculated based on the size and shape distributions of the NCs taken from Figure\,\ref{fig:temparam}a. (\textbf{d})  Direct measurement of TRM in Ga$\delta$FeN/Al$_{0.1}$Ga$_{0.9}$N after cooling down in a saturating $H=10$\,kOe and (\textbf{e})~its $T$--derivative. The dashed lines in (\textbf{d}) point to the superparamagnetic limit of about 500\,K.}
	\label{fig:MTRM1}
\end{figure}

The magnetothermal behavior of these ensembles of NCs traced for two orientations of $H$, {i.e.,} $H_\parallel$ parallel (full symbols) and  $H_\perp$ perpendicular (open symbols) to the film plane is shown in Figure\,\ref{fig:MTRM1}a and follows a trend specific to ferromagnetic nanoparticle ensembles previously reported for Fe-rich NCs stabilized in GaN \cite{Bonanni:2007_PRB,Navarro:2010_PRB,Bianco:2018_PCCP}. 
These features indicate that independently of the orientation, a specific distribution of energy barriers $E_B = K_{\mathrm{eff}} V_{\mathrm{NC}}$ for the ferromagnetic moment reversal determines the response in the whole temperature range. Here $K_{\mathrm{eff}} $ is the effective magnetic anisotropy energy density specific to a given NC with volume $V_{\mathrm{NC}}$.
The effect is particularly significant in the Ga$\delta$FeN/Al$_{0.1}$Ga$_{0.9}$N layer for $H_{\perp}$. 
This finding demonstrates that the predominantly prolate character of the $\varepsilon$-Fe$_3$N NCs in the layers grown on the Al$_x$Ga$_{1-x}$N buffers dramatically affects the magnetic anisotropy (MA), which will be treated in detail later. 

For an ensemble of non-interacting magnetic NCs the temperature derivative of the thermoremanence magnetization (TRM) provides qualitative information on the $E_B$ distribution in the ensemble\,\cite{Dormann:1997_ACP}. From $M_{\mathrm{TRM}} = M_{\mathrm{FC}} - M_{\mathrm{ZFC}}$, the  $-d(M_{\mathrm{FC}} - M_{\mathrm{ZFC}})dT$ is calculated and displayed in Figure\,\ref{fig:MTRM1}b, with non-zero values in the whole $T$-range and exhibiting a peak at around 50\,K. From this, the magnitude of the superparamagnetic limit $T_{\mathrm{SP}}$ in the layers is quantified. 
Here, $T_{\mathrm{SP}}$ is the temperature above which a given magnetic NC or an ensemble of NCs is in thermal equilibrium and is defined by $E_B=25 k_{\mathrm{B}}T_{\mathrm{SP}}$\,\cite{Bean:1959_JAP}, where $k_\mathrm{B}$ is the Boltzmann constant and the numerical factor 25 corresponds to the typical magnetometry probing time of 100\,s.

Due to the fact that all considered layers contain $\gamma$'-Ga$_y$Fe$_{4-y}$N NCs, their size distribution is taken into account. For each NC, the individual $K_{\mathrm{eff}} = K_{\mathrm{mcr}}  + K_{\mathrm{sh}} $, where $K_{\mathrm{mcr}}=\,(3 \times 10^5)$\,erg/cm$^3$ is the magnitude of the cubic magnetocrystalline anisotropy parameter of $\gamma$'-Fe$_{4}$N \cite{Coey:2008}, is calculated.
The positive sign indicates that the magnetic easy axes are directed along the $[100]$ direction, which is parallel to the $c$-axis of GaN.
The shape contribution to the MA for each NC:
\begin{equation}
K_{\mathrm{sh}} = (N_A - N_C )M_{\mathrm{sat}}^2/2\,,
\label{eq:Kshape}
\end{equation}

is determined by the difference $N_A - N_C$ of the demagnetizing coefficients $N$ of the considered nanocrystals according to the ellipsoid with semi-axes $A$ and $C$\,\cite{Osborn:1945_PR}.
The experimental magnitude of $M_{\mathrm{sat}} = 1700$\,emu/cm$^3$ established here is employed, considering that the main crystallographic axes of the NCs and their axes of revolution are aligned with those of the host lattice. 
The magnitudes of $K_{\mathrm{mcr}}$ and $K_{\mathrm{sh}}$ can be added with the $caveat$ that all NCs with negative values of $K_{\mathrm{eff}}$ 
are discarded. This is because for $K_{\mathrm{eff}} < 0$ the easy plane of the magnetization $M$ rotates smoothly by 180$^o$ to facilitate the reversal and the NCs are at thermal equilibrium at any $T$, thus not contributing to TRM.
Based on the data presented in Figure\,\ref{fig:temparam}a, as much as 50\% of the NCs belong to this category, a decisive factor for understanding the magnetic softness of the ensembles of NCs\,\cite{Bonanni:2007_PRB,Navarro:2010_PRB,Bianco:2018_PCCP,Navarro:2019_PRB,Gas:2019_MST}.
The large number of NCs in equilibrium explains also the low magnitude of $M_{\mathrm{FC}}$ (and $M_{\mathrm{TRM}}$), {i.e.,} less than 20\% of the total saturation value.
Finally, for nearly spherical NCs ($C/A \simeq 1$), where the cubic $K_{\mathrm{mcr}}$ prevails, $E_B = K_{\mathrm{eff}}V_{\mathrm{NC}} / 4$ is set, as expected for cubic anisotropy exhibiting magnetic easy axes oriented along the $\langle 100\rangle$ family of directions ($K_{\mathrm{mcr}}^{\mathrm{cubic}} > 0$)\,\cite{Walker:1993_JPCMa}. The calculated $T_{\mathrm{SP}}$ distribution as a function of the $K_\mathrm{eff}\,V_\mathrm{NC}/(25k_\mathrm{B})$ is depicted in Figure\,\ref{fig:MTRM1}c and is in agreement with the experimental data in Figure\,\ref{fig:MTRM1}b.
The calculated distribution peaks around 40\,K, decreases at higher temperatures, and remains non-zero up to 400\,K, as found experimentally.

The non-conventional behavior of $M_{\mathrm{ZFC}}$ and $M_{\mathrm{FC}}$ of the Ga$\delta$FeN/Al$_{0.1}$Ga$_{0.9}$N structure probed for $H_{\perp}$ indicates that even at $T\,=\,400$\,K the field of 100\,Oe is too weak to overcome the energy barriers. Therefore, direct TRM measurements to establish the actual magnitude of the low--$T$ $M_{\mathrm{TRM}}$ are performed. To this end, the sample is cooled down at a saturating field of 10\,kOe to $T\,=\,2$\,K, then the field is quenched and at $H\,\simeq\,0$ the TRM measurement is performed while warming up. For comparison, the same sequence is executed for $H_\mathrm{\parallel}$.
The results and their $T$-derivatives are presented in Figure\,\ref{fig:MTRM1}d,e, respectively.
The magnitude of the irreversible response increases for the perpendicular orientation (empty symbols) to about 80\% of the total  magnetic saturation.
Taking into account the significant MA of hexagonal $\varepsilon$-Fe$_3$N and the much weaker one of cubic $\gamma$'-Ga$_y$Fe$_{4-y}$N, the 80\% level is taken as a coarse estimate of the relative content of the $\varepsilon$-Fe$_3$N NCs in the layer grown on the Al$_{0.1}$Ga$_{0.9}$N buffer.

Both TRMs remain non-zero even at 400\,K. By extrapolating the curves to zero, with the maximum value of $T_{\mathrm{SP}}$ located at 500\,K. This procedure is valid because the derivatives $d M_{\mathrm{TRM}}/dT$ increase as $T \rightarrow 400$\,K. Interestingly, the $T$-derivative of $M_{\mathrm{TRM}}$ for the in-plane configuration is featureless and larger than the one established at low fields in the ZFC and FC measurements, suggesting that in these two measurements two different subsets of NCs determine the response.

\begin{figure}
	\centering
	\includegraphics[width=15 cm]{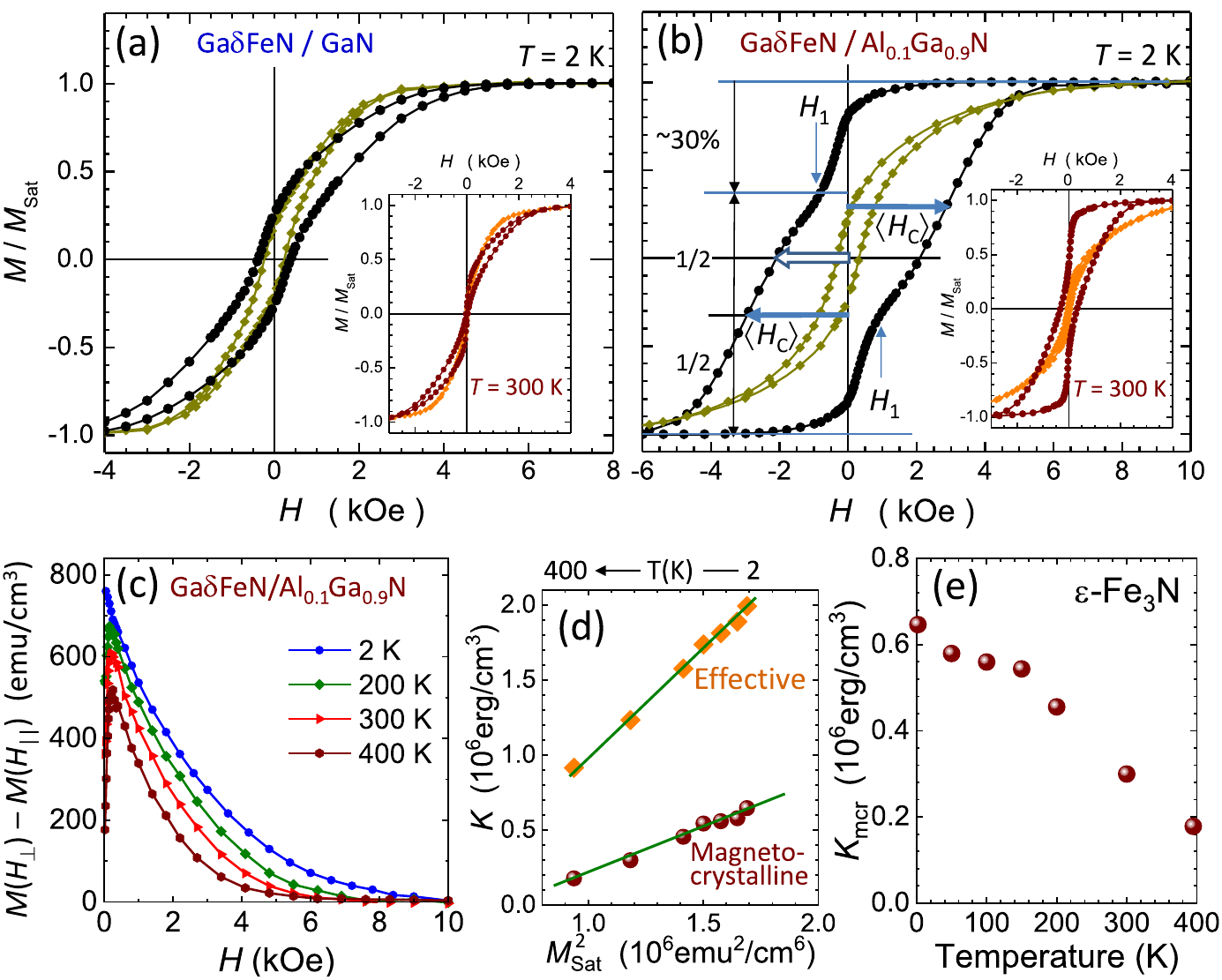}
	\caption{Normalized magnetization $M/M_\mathrm{sat}$ acquired at 2\,K for the two magnetic field configurations $H_\perp$ (circles) and $H_\parallel$ (diamonds) for (\textbf{a})  the reference sample, and (\textbf{b})  Ga$\delta$FeN/Al$_{0.1}$Ga$_{0.9}$N. The $M/M_\mathrm{sat}$ at 300\,K as a function of the magnetic field is depicted in the insets. The vertical arrows mark an inflection point $H_1$ on $M(H_{\perp})$ separating two different contributions to $M$ during its reversal. The~empty arrow marks the coercive field of the whole ensemble, whereas the lengths of the two full arrows indicate the average coercive field $\langle H_C \rangle$ of the prolate part of the distribution. (\textbf{c})  Magnetic anisotropy $M(H_{\perp}) - M(H_\mathrm{\parallel})$ obtained for the Ga$\delta$FeN/Al$_{0.1}$Ga$_{0.9}$N sample acquired at selected temperatures. (\textbf{d})~Magnitudes of $K_\mathrm{eff}$ established from the area under the curves in (\textbf{c}) plotted as the function of $M_{\mathrm{Sat}}^2$ (diamonds) and of $K_\mathrm{mcr}$ of $\varepsilon$-Fe$_3$N (bullets). Solid lines mark the proportionality of both $K_{\mathrm{eff}}$ and $K_{\mathrm{mcr}}$ to $M_{\mathrm{Sat}}^2$. (\textbf{e})  Temperature dependence of $K_\mathrm{mcr}$ of $\varepsilon$-Fe$_3$N.}
	\label{fig:M-H}
\end{figure}
The normalized magnetization $M/M_\mathrm{sat}$ of the layers as a function of the magnetic field is presented in Figure\,\ref{fig:M-H}a,b, where both $M(H_{\perp})$ and $M(H_\mathrm{\parallel})$ show the sensitivity of the magnetization to the orientation of $H$ for the reference structure and for the Ga$\delta$FeN/Al$_{0.1}$Ga$_{0.9}$N layer, respectively. 
The measured $M(H)$ saturates beyond $\pm\,10$\,kOe and does not significantly depend on $H$ in the whole studied $T$-range, as demonstrated earlier in Figure\,\ref{fig:M1}a for the reference sample and in previous studies~\cite{Navarro:2019_PRB,Gas:2019_MST}. 
A similar behavior is observed for all the layers deposited on the Al$_x$Ga$_{1-x}$N buffers. 

It is worth underlining that the main symmetry axes of the $\varepsilon$-Fe$_3$N NCs are fixed in the direction of the $c$-axis of GaN, {i.e.,} perpendicular to the sample plane, which is essential for modelling the results. The uniaxial magnetocrystalline anisotropy (UMA) of the hexagonal $\varepsilon$-Fe$_3$N NCs was found to be between (0.5--1$\times 10^6$)\,emu/cm$^3$ \cite{Mamiya:2002_TMSJ} with the easy axis along the $[0001]$-direction. Due to preferential nucleation along the dislocations, the distribution of shapes of the $\varepsilon$-Fe$_3$N NCs is highly asymmetric, adding a sizeable shape contribution to the native crystalline UMA of $\varepsilon$-Fe$_3$N. 
The data presented in Figure\,\ref{fig:temparam}c yield the average elongation $\langle C/A \rangle = 1.34$ for the prolate part of the distribution, what, according to Eq.\,\ref{eq:Kshape} and $M_{\mathrm{sat}}\,=\,1300$\,emu/cm$^3$, points to $\langle K_{\mathrm{sh}}\rangle\,=\,(1.2 \times\,10^6)$\,erg/cm$^3$, which represents the most relevant contribution to the overall MA of this ensemble.

The large UMA along the growth direction is the origin of the pronounced squareness and the resemblance of the experimental $m(H_{\perp})$ to the perpendicular magnetic anisotropy of bulk ferromagnets and layered structures. This is further demonstrated by the hard-axis-like shape of $m(H_\mathrm{\parallel})$.
The magnitude of the UMA exerted by the considered ensemble of NCs is calculated by taking the experimental difference $\Delta M(H)\,=\,M(H_{\perp})\,-\,M(H_\mathrm{\parallel})$, plotted for selected temperatures in Figure\,\ref{fig:M-H}c.
By definition, the area under the $\Delta m(H)$ yields the magnitude of $K_{\mathrm{eff}}$. 
The established magnitudes are plotted against the corresponding magnitudes of $M_{\mathrm{sat}}^2$ in Figure\,\ref{fig:M-H}d (diamonds). 
The nearly linear relationship $K_{\mathrm{eff}} \propto M_{\mathrm{sat}}^2$ confirms the significant UMA in this ensemble, allowing the direct determination from Eq.\,\ref{eq:Kshape} of $K_{\mathrm{mcr}}$ of $\varepsilon$-Fe$_3$N from the $T$--dependence of $m_{\mathrm{NC}}^{\mathrm{sat}}(T)$ (Figure\,\ref{fig:MTemp}). 
The resulting magnitudes of $K_{\mathrm{mcr}} = K_{\mathrm{eff}} - K_{\mathrm{sh}}$ established at all the measured temperatures, are shown in Figure\,\ref{fig:M-H}e (bullets). This is the first direct determination of the absolute magnitudes of $K_{\mathrm{mcr}}$ of $\varepsilon$-Fe$_3$N in such a broad and technologically relevant temperature range up to 400\,K.

On the other hand, as indicated in Figure\,\ref{fig:M-H}b, the magnetization process in the Ga$\delta$FeN/Al$_x$Ga$_{1-x}$N structures is based on two rather independent switching processes. This is seen at the two temperatures exemplified in Figure\,\ref{fig:M-H}b. The $T\,=\,2$\,K case, where the thermal activation contribution to $m(H)$ can be neglected, is considered in detail. Here, about a third of the total magnetization of the NCs switches at very weak fields. This process completes at weak negative fields, where a kink is seen in $m(H_{\perp})$ at about $\pm 1$\,kOe, marked by the arrows at $H_1$.
Up to $H_1$ about 30\% of the total $M$ has switched or rotated to the new direction of $H$.
This is the result of a narrow band of weak switching fields brought about by the minority of the oblate NCs (which nominally reverse $M$ at $H = 0$) and of several cubic $\gamma$'-Ga$_y$Fe$_{4-y}$N NCs, which reverse $M$ at weak fields, as demonstrated in Figure~\ref{fig:M-H}a. For the remaining 70\% NCs, the switching process begins after $H_{\perp}$ passes $H_1$ and these are the prolate $\varepsilon$-Fe$_3$N NCs, which, due to their generally high $K_{\mathrm{eff}}$ require larger magnitudes of $H$ to overcome the individual anisotropy fields $H_A = 2 K_{\mathrm{eff}} / M_{\mathrm{sat}}$.
Since the majority of the NCs is in the single domain state, the different magnitudes of $H_A$ contribute to a broad distribution of switching (coercive) fields $H_\mathrm{C}$, resulting in the  wide $m(H_{\perp})$ for $|H| > |H_1|$.
From the magnitude of $\langle K_{\mathrm{eff}} \rangle$, $\langle H_C \rangle = 3$\,kOe at low temperatures is obtained and it is also extrapolated directly from the $m(H)$ curve in Figure\,\ref{fig:M-H}b. 
Since the reversal process of $M$ of the prolate fraction of the NCs ensemble in the Ga$\delta$FeN/Al$_x$Ga$_{1-x}$N structures starts after the magnetically soft part of the ensemble has reversed, the $H_\mathrm{C}$ cannot be determined at $M = 0$.
The $m(H)$ after $H_1$ is assigned to the prolate $\varepsilon$-Fe$_3$N, marked by the arrows in Figure\,\ref{fig:M-H}b, from where the corresponding $\langle H_\mathrm{C} \rangle$ can be obtained.
It is worth noting that the difference in  $\langle H_\mathrm{C} \rangle$ between the two branches of $m(H_{\perp})$ corresponds to the magnitude of the soft part of $M$ which switches within $|H| < |H_1|$, \textit{i.e.} the magnetically hard part of $m(H_{\perp})$ corresponding to the prolate NCs is broken up by the magnetically soft component of the distribution.

 \section{Conclusions}
Strained and partially relaxed Ga$\delta$FeN thin layers grown on Al$_x$Ga$_{1-x}$N buffers by MOVPE reveal the formation of hexagonal $\varepsilon$-Fe$_3$N and fcc $\gamma$'-Ga$_y$Fe$_{4-y}$N nanocrystals epitaxially embedded in the GaN matrix. The Ga$\delta$FeN layers are strained for an Al concentration in the buffer up to 10\% and then relax up to 85\% for an Al concentration of 41\%. With increasing Al content, an increase in the dislocation density in the buffer layers is observed, together with a preferential aggregation of nanocrystals along the dislocations in the Ga$\delta$FeN layers. The NCs have either oblate or prolate shape, with the majority of the NCs being prolate. Both nanocrystal phases are coherently embedded into the surrounding GaN matrix with an epitaxial relation:  $[0001]_\mathrm{NC}\parallel[0001]_\mathrm{GaN}$ and $\langle11\bar{2}0\rangle_\mathrm{NC}\parallel\langle10\bar{1}0\rangle_\mathrm{GaN}$ for the $\varepsilon$-Fe$_3$N NCs, and $[001]_\mathrm{NC}\parallel[0001]_\mathrm{GaN}$ and $\langle110\rangle_\mathrm{NC}\parallel\langle11\bar{2}0\rangle_\mathrm{GaN}$ for the $\gamma$'-Ga$_y$Fe$_{4-y}$N NCs. 

The magnetic response of the layers is consistent with the one previously found for phase-separated (Ga,Fe)N consisting of two components: a dominant paramagnetic low-$T$ contribution from Fe$^{3+}$ ions dilute in the GaN matrix and in the buffer volume, and a ferromagnetic one dominant above 50\,K originating from the $\gamma$'-Ga$_y$Fe$_{4-y}$N and the $\varepsilon$-Fe$_3$N embedded NCs\,\cite{Bonanni:2007_PRB, Navarro:2010_PRB}. The low--$T$ contribution of the Fe$^{3+}$ ions to the total magnetization reaches magnitudes comparable to those of the NCs. The $T_\mathrm{C}$ of the reference layer containing solely $\gamma$'-Ga$_y$Fe$_{4-y}$N is found to be $(630\,\pm\,30)$\,K, pointing at the inclusion of Ga into the NCs and therefore lowering the $T_\mathrm{C}$ with respect to one of $\gamma$'-Fe$_4$N\,\cite{Shirane:1962_PR}. Due to the formation of additional $\varepsilon$-Fe$_3$N in the Ga$\delta$FeN/Al$_x$Ga$_{1-x}$N layers, $T_\mathrm{C}$ is increased to $(670\,\pm\,30)$\,K, indicating a high crystalline and chemical quality of the NCs. Moreover, the calculated magnetization of the NCs is consistent with literature values. The magnetization process in the Ga$\delta$FeN/Al$_x$Ga$_{1-x}$N structures is based on two substantially independent switching processes: a~relatively fast switching of the oblate and $\gamma$'-Ga$_y$Fe$_{4-y}$N NCs at low fields, followed by the switching of the $\varepsilon$-Fe$_3$N NCs, which require larger magnitudes of $H$ to overcome the individual anisotropy fields. All Ga$\delta$FeN layers grown on the Al$_x$Ga$_{1-x}$N buffers exhibit a sizeable uniaxial magnetic anisotropy with the easy axis matching the $c$-axis of the hexagonal $\varepsilon$-Fe$_3$N NCs and the $[0001]$ growth direction of the layers. This suggests that the formation of ordered elongated hexagonal $\varepsilon$-Fe$_3$N NCs along the dislocations 
in the  Al$_x$Ga$_{1-x}$N buffers is responsible for the observed out-of-plane magnetic anisotropy. The finding is substantiated by the value of $H_\mathrm{C}$ obtained directly from the normalized magnetization for $H_\perp$ that is well reproduced by the calculated value obtained considering the  $K_{\mathrm{eff}}$ of the prolate $\varepsilon$-Fe$_3$N NCs. Significantly, this is the first direct determination of the absolute magnitudes of $K_{\mathrm{mcr}}$ of $\varepsilon$-Fe$_3$N in a broad and technologically relevant temperature range up to 400\,K. 

According to these findings, Ga$\delta$FeN/Al$_x$Ga$_{1-x}$N heterostructures provide a controllable housing for stabilizing ordered arrays of ferromagnetic Fe$_y$N compounds, opening wide perspectives for spin injection in these phase-separated material systems and for the electric-field manipulation of the magnetization\,\cite{Sztenkiel:2016_NComm}.\\

\begin{acknowledgments}
The work has been funded by the Austrian Science Fund FWF Projects No. V478-N36, P26830 and P31423, and the Austrian Exchange Service (\"{O}AD) Project No. PL-01/2017 (DWM.WKE.183.72.2017). The financial support by the Austrian Federal Ministry for Digital and Economic Affairs, the National Foundation for Research, Technology and Development and the Christian Doppler Research Association is gratefully acknowledged. The authors greatly acknowledge Werner Ginzinger for his extensive work in the sample preparation and on TEM measurements. 
\end{acknowledgments}



\end{document}